# Use and Usability in a Digital Library Search System

*Robert K. France, Lucy Terry Nowell*, Edward A. Fox, Rani A. Saad, and Jianxin Zhao*

Virginia Tech Digital Library Research Laboratory
840 University City Boulevard
Blacksburg, VA 24060 USA
E-mail: france@vt.edu, fox@vt.edu, rsaad@vt.edu, jxzhao@csgrad.cs.vt.edu

*Battelle Pacific Northwest Division
902 Battelle Boulevard
P.O. Box 999
Richland WA 99352 USA
E-mail: Lucy_Nowell@pnl.gov

## ABSTRACT

Digital libraries must reach out to users from all walks of life, serving information needs at all levels. To do this, they must attain high standards of usability over an extremely broad audience. This paper details the evolution of one important digital library component as it has grown in functionality and usefulness over several years of use by a live, unrestricted community. Central to its evolution have been **user studies,** analysis of **use** patterns, and formative **usability evaluation**. We extrapolate that all three components are necessary in the production of successful digital library systems.

## INTRODUCTION

MARIAN is an innovative search system suitable for digital libraries. For the past six years it has served as an alternative catalog for the main Virginia Tech library, providing remote free-text search capabilities to the Tech community. The history of MARIAN serves as an instructive example of the evolution of a key digital library component in the hothouse environment of the World Wide Web. We present MARIAN's transition from an abstract conception in the minds of information retrieval specialists to a useful and pleasant system. It is our contention that such a transition can only be accomplished through an emphasis on users, use, and usability.

**Users** are the *raison d'être* for any computer system, Digital library users are a unique community of information seekers. On the one hand, they can be presumed to have serious and sophisticated information needs, although they are sometimes unsophisticated in expressing those needs. On the other hand, library users have been aptly characterized as "chronic beginners." This is in large part due to their uneven demand for library services, often characterized by periods of intense research separated by long gaps. The most sophisticated query expression skills can easily be lost when they are not exercised. In general, library systems must cope with a wide range of users, from the casual to the compulsive, and a wide range of information needs, from general to specific.

User studies have been part of library and information science for at least a century. Studies of online library catalog users [2, 6, 9, 10] provided important corrections to the design of such systems, including MARIAN. Much more remains to be done [7]. Recent studies of readers and writers [1, 15, 19] are important to digital library design, as are studies of interactions with intermediaries [22]. Crucial to the design of integrated digital library systems are integrated studies of library users' behavior [e.g., 4, 5, 16, 21]. These have proven difficult even in physical libraries. Studies of digital libraries, where most transactions occur over the network, will prove even more challenging [14]. It is vital that such studies continue.

Patterns of **use** can help us understand how well our systems serve their community. They can help us guide system evolution both on the small scale of system tuning and debugging and on the large scale of feature and system development. Later in this paper we describe use studies of MARIAN based on semi-automatic log analysis and how the data gained therein have guided changes to the system.

System change can also be guided by more formal **usability** studies. Our third section describes formative usability evaluation performed while MARIAN was still being developed that resulted in several profound changes in system presentation and functionality. It is the combination of formal usability testing, statistical analysis of system use, and in-depth study of our user community that has produced a system as well-subscribed as MARIAN.

## MARIAN INITIAL DESIGN

MARIAN grew out of a perceived need for improved search functionality rather than good human-computer interaction. Specifically, the system specifications (see Fig. 1) revolve around the mechanics of search and retrieval. They are a direct reaction to common critiques of the library catalog systems of their day (e.g., [6]) – those who remember these will notice that the "dread zero-item search" is addressed explicitly – and to our own and others' observations of library users' behavior. Although the issues addressed, from ease of use to serendipity, are framed in user-oriented terms, the solutions are presented in terms of data representation and matching functions.

> "(1) The system should be easy to use with no training, but with sophisticated control available. (2) Every search should produce something. In particular, where no records exactly match the query, partial marches should be shown to the user. (3) No search should overwhelm the user, even those with many matching records. (4) The system should be extensible, reliable, fast, and *networked*. (5) Finally, the system should support serendipity It should support results presented in a context, close matches *as well as* exact matches, and browsing. Some unpredictability is acceptable as a consequence of achieving this goal, but repeatability must never be sacrificed."
>
> "Treat words as words. Treat names and subjects as individuals. Treat catalog records as collections of features. In all cases, provide a partial match function."

**Figure 1: Excerpts from MARIAN specification and design documents.**

Another influence was an earlier study [11] at Virginia Tech where different search engines running on a common database of library catalog records were compared under a common user interface. Users were set the task of assembling "good" lists of sources for a variety of topics. To a significant degree they preferred approximate matching systems (vector search with or without feedback) to constructed-key or Boolean keyword search engines. Users found the approximate match engines easier to use, and they were able to reach closure on their "good" lists in shorter time. Thus MARIAN was designed around approximate match searching.

A central feature of the MARIAN design is the use of *multiple access points.* People commonly remember books by descriptions that cross categories of library data. "Conway's book on number theory," for instance, contains a partial description of an author – *Somebody* Conway – and a text phrase that might occur in either title or subject field. MARIAN was designed to make such queries easy.

Finally, MARIAN was designed to promote query re-use and refinement. We believe that library users are unlikely to be satisfied with a single interaction with the catalog system. This is particularly true in the context of a digital library, where additional system interaction must take the place of using a different catalog or roaming the stacks. Thus MARIAN supports a query history that encourages users to revisit old searches in any order, easy means for query resubmission and editing, cut-and-paste facilities that promote using parts of catalog records as new queries, and (in design, though not in any released version to date) a simple mechanism to use catalog records themselves as queries. We intended exploration through query modification to be the user's basic interaction mode with MARIAN.

The dialog between a user and MARIAN takes the form of an exchange of information objects, some of which are simple and some quite complex. For instance, a key sequence in the bibliographic search task consists of MARIAN sending the user a bibliographic query form, the user replying with a specific query, and MARIAN replying with a set of results. Each information object includes both data and functionality: for objects like the query form and result set, much of that functionality is bound up with user actions.

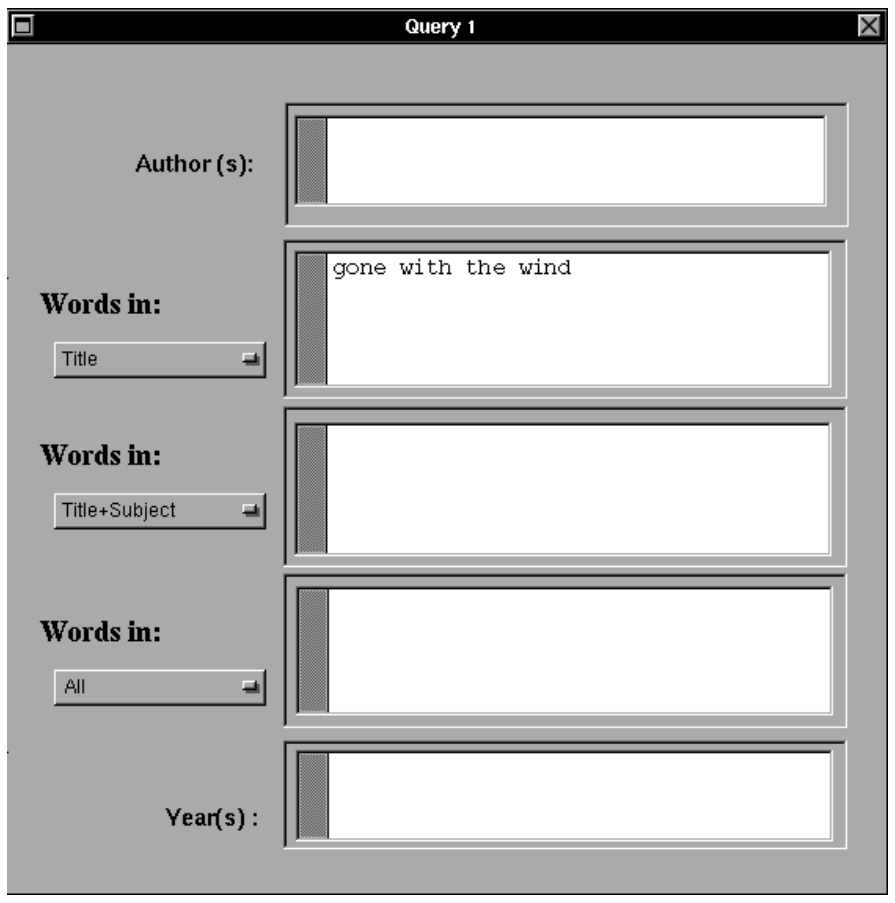

**Figure 2: First implementation of the query form object as a window.**

In the first version implemented, each information object was presented to the user as a window, using the paradigm exemplified by NeXTStep, Macintosh, and X-Windows systems. Fig. 2 shows the original presentation of the query form object, and Fig. 3 the result set object. The query form in this version includes five fields, providing up to five simultaneous access points. The top and bottom are fixed to "Author" and "Date" respectively, but the three in the middle are configurable by pop-up menus, shown "popped" in the middle field of Fig. 7 below.

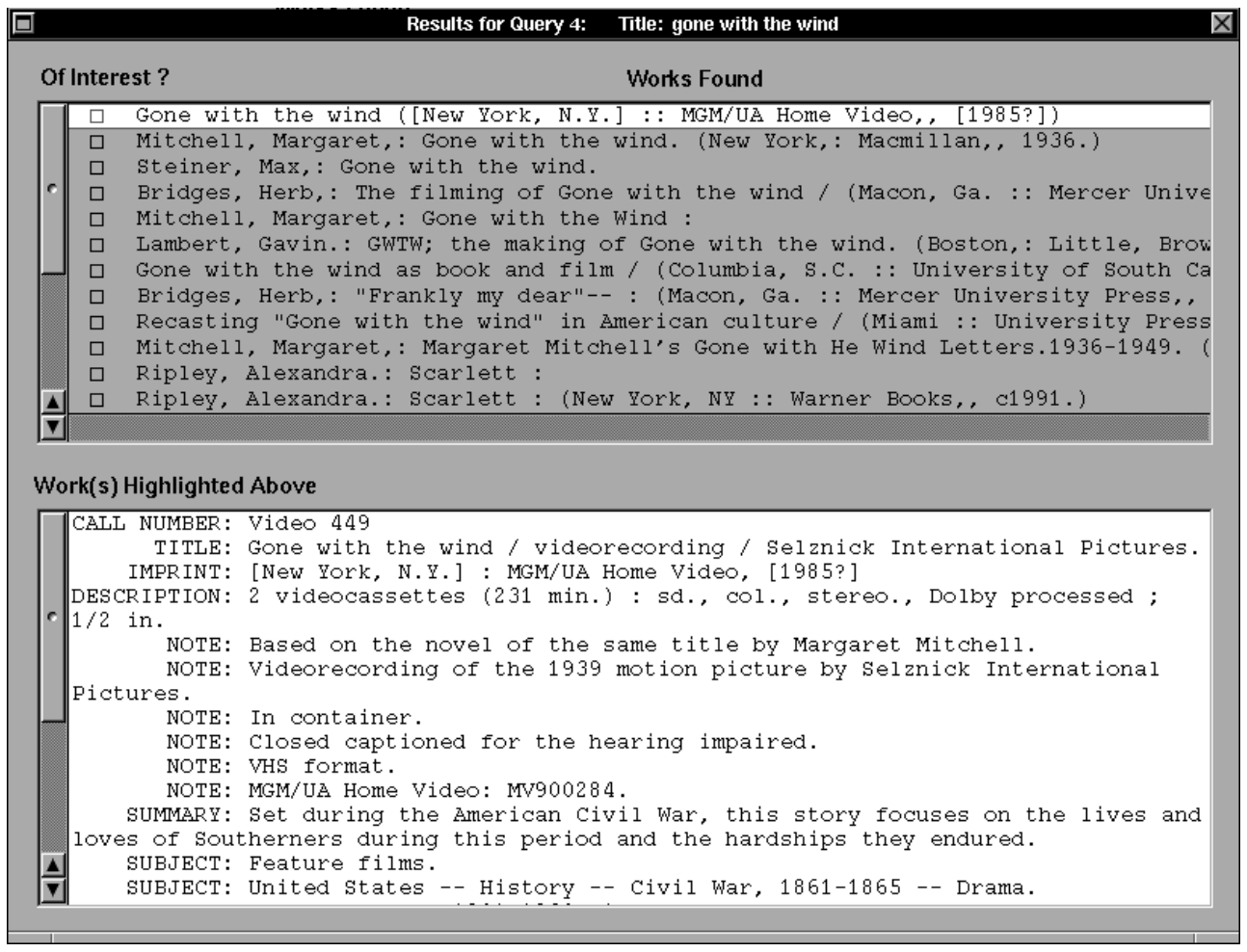

**Figure 3: First window implementation of the result set object.**

As with other ranked retrieval systems (e.g., Okapi [20]) MARIAN often presents result sets for which the user may only be interested in the top 1/2 of one percent. The result set object shows the top segment of the full result set, with the option of then showing more results. Results highlighted in the upper pane of the window are shown in detail in the lower pane. Other user actions for this object include *marking* items as being of interest, and then either saving or printing the marked set or using it as the seed for a "feedback" search for similar items.

As in the Macintosh paradigm, a set of cascading menus (not shown) is also present in NeXTStep interfaces, and in this version of the MARIAN interface some user actions for each object were located in the menus. In particular, both the "Show more results" and the "Find more like (a group of results)" functions for the result set object were options under the "Search" menu, as was the query form action of dispatching a completed query to the system.

## USABILITY TESTING

MARIAN is implemented as a network of central server modules which handle search and collection management and a set of clients that manage user interface presentation. From design time through the first production version, each client was presumed to manage a single session, running either on a public terminal or on a user's desktop machine. In 1992, with minimal functionality completed both in the back end and in a remote client for NeXTStep computers, and with a small but realistic collection loaded, we made a formative usability evaluation of the system.

Following current practices [13], we collected detailed data from extended sessions with a small sample of users. Participants were selected to be representative of the intended user community, including students, faculty, clerical staff and a reference librarian. Our intent was to cover a wide range of abilities and sophistication, not to constitute a statistically balanced sample. This is in keeping with our purpose in formative evaluation: to uncover problems in using the system, not to approximate how the system would actually be used.

Each observation session had three parts. First, the participant was introduced to the evaluators and the process, and generally put at ease that it was the system that was being evaluated, not he or she. Participants not used to the NeXTStep interface, to windowing systems, or to a mouse (yes, there were some of these) were given all necessary instruction. No instructions were given for MARIAN, however, since we needed to discover whether the system was usable by the "chronic beginner."

System designers worked with a usability evaluator to design a set of benchmark tasks for MARIAN (see Fig. 4). These tasks were then broken down into steps to be followed by a participant. At the end of the session each participant completed a questionnaire, evaluating MARIAN on several scales. The questionnaire elicited further data on what was usable and what less so, and included questions designed to illuminate the users' mental models of the underlying search process.

**Benchmark Tasks for MARIAN Usability Testing**
**Sept. 16, 1992**

4. Without creating a new information request:

   a.Try to find and say aloud the author and call number of *Men under Water.*

   b.Identify three other works that MARIAN found in response to your request. Say aloud the titles and call numbers.

   c.Why do you think these works were found?

5. How might you have MARIAN attempt to find more works that are similar to *Man under the Sea* and *Men under Water* ? [This feature does not work yet. Please just tell us how you would try to do this task.]

...

7. Create a new request for information, asking MARIAN to locate works whose titles contain the words "women" and "American history." Do not have MARIAN actually locate the works.

8. a.Change the request just created, so that the words "women" and "American history" may be anywhere in the works.

   b.Have MARIAN locate the works.

9. a.How many works did MARIAN find?

   b.What difference, if any, do you see between the first works listed and the last few?

   c.Have MARIAN try to locate more works that contain the words "women" and "American history."

   d.How do these seem to differ from the works already found?

**Figure 4: Excerpts from benchmark task instructions.**

Evaluation sessions were conducted by two people, one to lead the participant through the tasks and to act as a help system (the integrated MARIAN help system being unfinished at the time) and another, trained in usability evaluation, to note the participants' behavior. Particular attention was given to **verbal protocols** – statements by the participants – and **critical incidents** – moments of insight or criticism, good or bad.

After all sessions had been conducted, task completion evaluated, and verbal protocols and critical incidents collected and transcribed, all observations were broken up and shuffled, then regrouped by similarity. As hoped, this process permitted the formation of emergent categories, which were then used to organize the full set of results (see Fig. 5). A programming team evaluated each result and estimated the cost of implementing the change. We combined the two into a set of recommendations for changes, with priorities set by how easily each change could be made and how much it would improve usability (Fig. 6).

**Recommendations Based on MARIAN Usability Testing**
Lucy Nowell, nowell@vtcc1
October 13, 1992

1. Revise menus/menu structure
   - Restructure to reflect task organization, not software organization
     - SEARCH
       - CREATE QUERY or NEW QUERY
         - CHOOSE SOURCE
       - DO SEARCH
       - FIND MORE LIKE...
       - MODIFY QUERY

2. Provide feedback
   - MARIAN started
   - Search status
   - System failure

3. Communicate principles of operation
   - Matching principle(s) for queries
   - Fixed retrieval set size
     - Availability of more documents
   - Ordering principle in results

4. Put the user in control
   - Maintain user selections (highlighted items) after feedback search
   - Maintain query number until query is modified
   - Maintain user settings
     - Window size(s)
     - Results order/format

5. Make the user's task easier
   - Eliminate case sensitivity
   - "Do Search" button
   - "Equal opportunity interactions"
   - User control of ordering / layout of results
     - Rank order vs. alphabetical by author
     - Columns vs. text scroll
   - Access to query while viewing results
     - Without changing query number

6. Use consistent terminology
   - Menu text with related window names

**Figure 5: Overview of goals reflects the taxonomy of usability problems.**

**I. Immediate changes (high utility, low cost)**

…

4. Change "All" to "Any part of description" in coverage pop-up menu
   Change ampersands to pluses.

5. Add some explanatory text to BiblioQueryForm object (see example in attached figure).

6. Add **Perform search** button to BiblioQueryForm window
   Keep item in **Continue search** menu and command key shortcut

…

**II. Near-term (v 1.1 - 2.0) changes (lower utility &/or higher cost)**

1. Change pop-up coverage menu in Bibliographic Query Form window to pop-up modal object as illustrated in attached.

2. Add "exact match" capability to all searchers.
   Test a sample of users to find out whether "exact match" means:
   All words given in order given with no additions
   All words given in order given, additional words (or additional function words) OK
   All words given in any order, additional words (or additional function words) OK

…

9. Separate check-boxes in NeXTStep RetrSetDisplay from text by more white space
   Provide label: "Double-click box to indicate interest" in small font.

10. Analyze MARC data with no attention to the case of words. Test whether this works better.

**III. Long-term (post v. 2.0) changes (low utility &/or very high cost)**

1. Add a persistent, but closable **Status** window to dynamically report the state of the session and the progress of active queries.

2. Add extended Boolean capability, together with Boolean operators in and among query fields

3. Add context-sensitive help to interfaces that support it.

4. Allow users to choose (as a **Preference**) from among types of BiblioFormQuery objects

**Figure 6: Excerpts from change schedule arising from the usability evaluation.**

The formative usability evaluation uncovered usability problems with both query form and result set objects. Most striking, perhaps, were the difficulties users had integrating actions located in menus and windows. For instance, users often got badly stuck when they had filled out a query form object but could not find how to submit it for searching – a critical incident of the worst sort. Similarly, it was clear from verbal protocols that without prompting from the task steps, most users would not even have discovered the "Find more like" action in the result set object. As a consequence, both windows were redesigned to include buttons implementing these actions (Figs. 7 and 8).

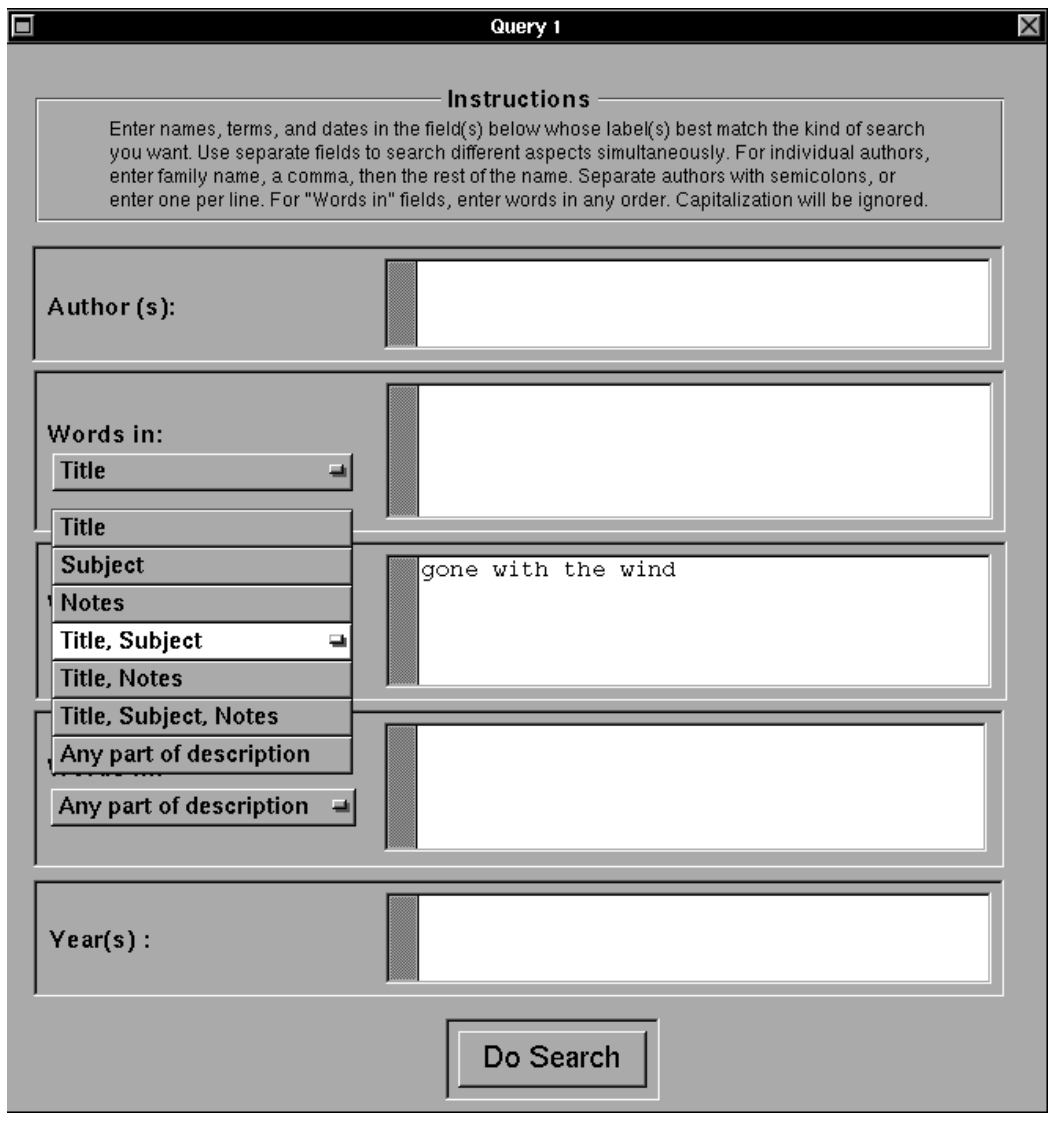

**Figure 7: The query form object as revised after formative usability evaluation.**

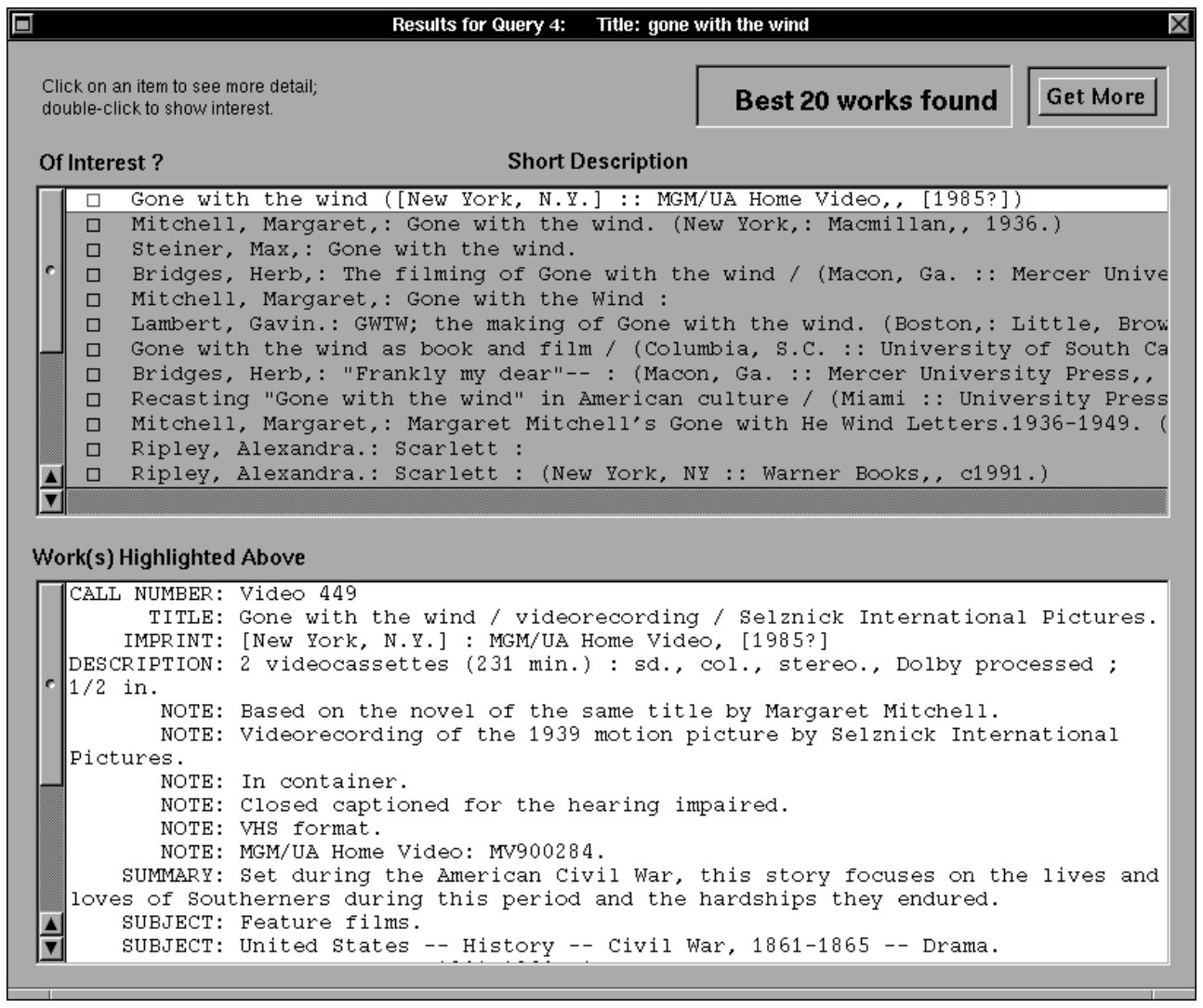

**Figure 8: The result set object after usability evaluation.**

Severe usability problems were apparent in the action of extending the result set. Typical verbal protocols are shown in Fig. 9. Several modifications were made to this feature: The number of results in the segment was displayed prominently on the results set window, together with the button for extending the set (compare Fig. 3 and Fig. 8). Horizontal lines were used to separate batches as an aid to navigation, and the window functionality was changed so that highlighted items were maintained when new batches were added.

> I have no clue. Can't tell what's old or new, or if there is anything new.
>
> I don't like this endless scroll of text. [Kept expanding set, 3 or 4 times.] I have no indication that it's working...
>
> [chose to find more from query] - I have no idea if it's working. I'm confused - I had 6 things selected. It should not have unselected those. They should still be here, with other added at the end of this. Irritated me. I wanted to compare to what I had selected.
>
> [Counting found docs.] Don't know if it tells me somewhere on the screen. I would like to know how many, how far I would have to scroll down.
>
> Same window, but it has more things here. I can't tell...but I would have looked through the last... I expected a second window: Q2.1.2, or something. But a line or something in the window to help me know where the new stuff starts would help.

**Figure 9: Some verbal protocols collected while study participants were trying to expand a result set.**

Usability testing also provided positive feedback. Verbal protocols indicated that users were at first confused by the query history, but rapidly accepted it as a useful tool. We also received further confirmation that users appreciate approximate search as long as they can model the match criteria ("Tries what I asked for first and then gives me more options in results. That's the way I think." and "Starts with exactly what I [asked for] and then keeps branching, if it didn't give what I want. I like that it's doing that for me...") and that the system does indeed promote a certain amount of serendipity ("Look at all I found ... I like to pore through it all ...").

The value of the usability study exceeds the individual changes that it motivated. It was also the vehicle by which MARIAN's interaction with users became more consistent, from font usage to consistency of action. It conclusively resolved an ongoing discussion among the designers as to the value of expending window "real estate" on instruction text. Instruction text was clearly desirable, and was added to both query form and result set windows.

On a different note, the observation sessions themselves were valuable to the system designers. Prior to the evaluation, several key designers had been openly doubtful of the value of usability testing. Even these needed only to serve as observers in a single session before they recognized the importance of the process. There is nothing that carries quite the weight of seeing a user struggling to find or use a system feature that you were *sure* was intuitively obvious. Watching users interacting with the system also gave all involved a better feel for the model that users form of the system, and thus of how to design features with that model in mind.

## INTO THE WEB

MARIAN was first released to the Virginia Tech campus in 1993. At that point the system included the NeXTStep interface, modified in keeping with the usability tests, an ASCII interface for text terminals, and a "batch" mode that could be run through a gopher script. X-Windows and Macintosh interfaces were under construction when the explosion of the World-Wide Web passed through Tech, changing the face of remote interaction here as everywhere.

When it became clear that Web browsers were here to stay, the MARIAN development group abandoned work on platform-specific clients and switched to producing a Web gateway ("WebGate"). Our goals for Web interaction with the MARIAN back end were to provide as nearly as possible the power and freedom of the standalone clients. Specifically, we wanted to keep the notion of a session with query and result histories, of query modification and reuse, of result set extension and subsetting, and of mixed initiative [11]. Few of these goals were met in the first generation Web gateway.

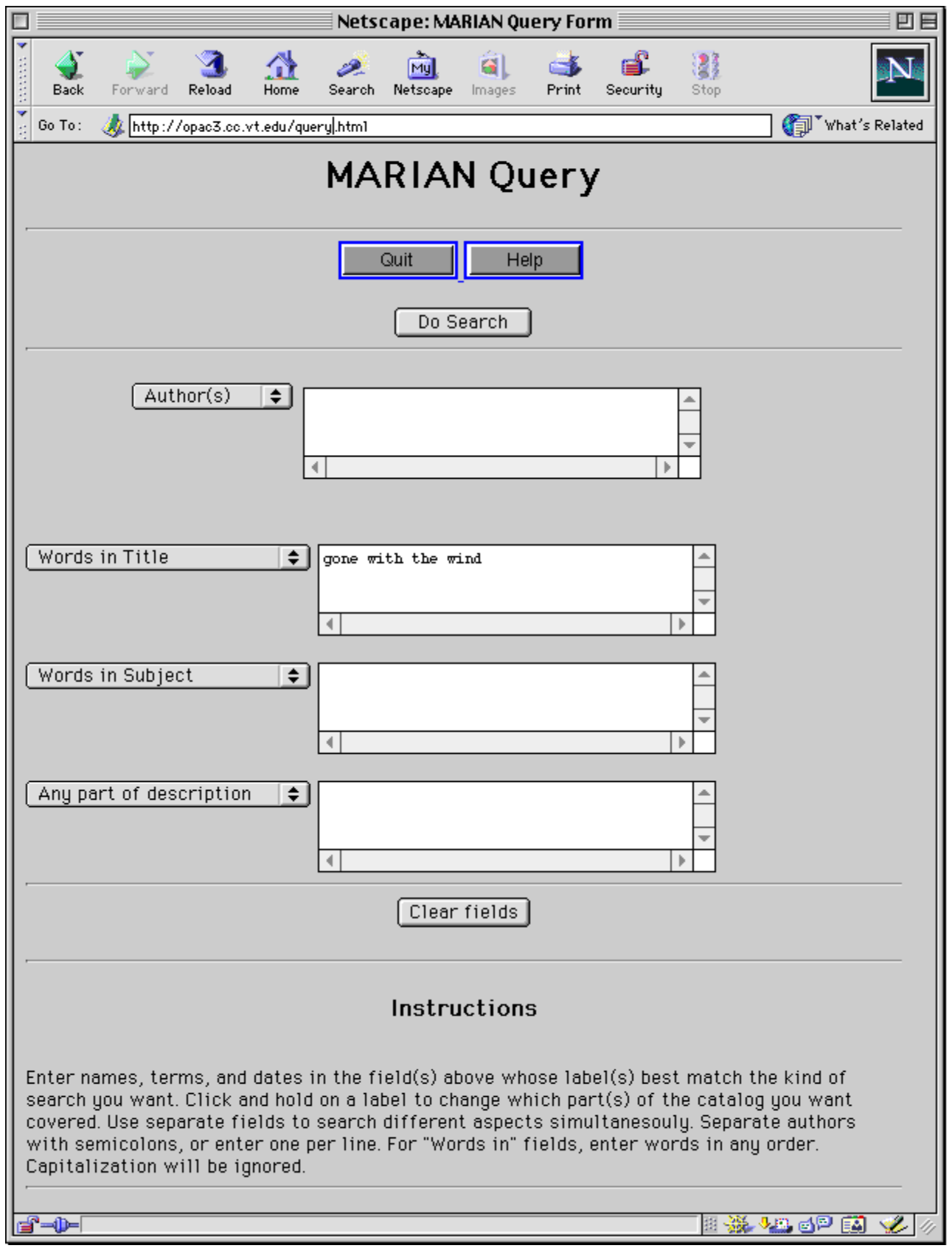

**Figure 10: Initial Web page implementation of the query form object, shown in a very tall browser window.**

We also wanted to maximize the usability of the Web interaction. Specifically, we wanted pages to be fast loading but pleasant to view, we wanted users to be able to accomplish the core tasks with minimal keystrokes, minimal mouse clicks and especially minimal page transitions, and we wanted users to be able to navigate within MARIAN using only buttons on the pages, but also to have the system behave well when users used the “Back” or “Go” functions of their browsers. These goals were met.

All functions that had been part of the persistent menu structure in the windowing interfaces were converted to buttons in the Web pages. This included both context-free actions such as starting a new query or quitting a session, and context-sensitive actions such as obtaining help on the current page. A set of state variables, embedded in dynamically generated URLs, made it possible to subvert HTTP’s state-less transaction protocol. This was necessary to differentiate old queries and result sets from new ones and thus to allow query re-use and result set extension.

Converting the query form object to HTML occasioned little difficulty as the pop-ups, buttons and text fields used in our windowing interfaces already existed in the HTML “form” construct, as did the actions of clearing fields and submitting the form (Fig. 10). Instructions were moved to the bottom of the form so that as much of the active fields as possible would be immediately visible to the user.

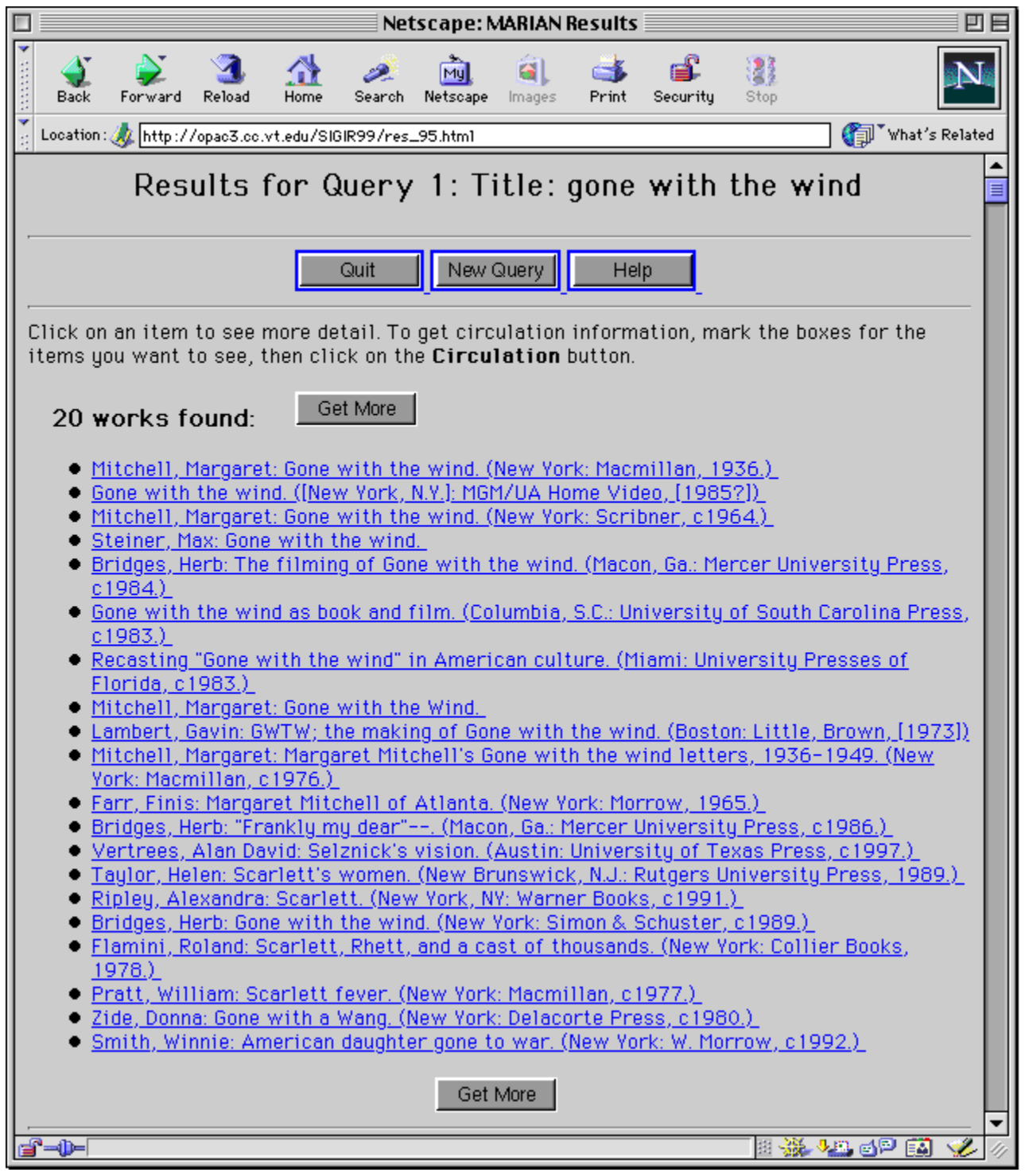

**Figure 11: Initial Web page implementation of the result set object. Full bibliographic record presentations are now a hyperlink away from the result list.**

Converting the result set object required more radical changes. Working within the confines of a browser window, we could not manage the two linked panes that we could in NeXTStep, X-Windows or Macintosh environments. Instead, we chose to hold the full bibliographic descriptions in WebGate, linking each to its short description with a hyperlink (Fig. 11).

## USER LOG ANALYSIS

As part of normal system operation, MARIAN logs performance of all back end modules. These logs serve as debugging aids and performance measures. Every effort is made to ensure confidentiality and to prevent any session record from being associated with specific users. In 1996 another use was found for these logs, as logs from summer and fall were collected and analyzed for user behavior patterns. We studied timings and analyzed user interactions with result sets and documents, but the conclusions that had most effect on the interface were those related to queries.

MARIAN sessions in the period studied ranged from a single query to a dozen, falling off in a typical Zipfian curve (Fig 12) (see [3] for a discussion of the prevalence of this sort of curve). More than a third of the sessions included multiple queries, which validated our assumption that query refinement is an important user activity, and permitted some analysis of how human queries evolve. It also motivated us to reinstate query editing to the Web interface and to advance reinstating the query history to a higher priority.

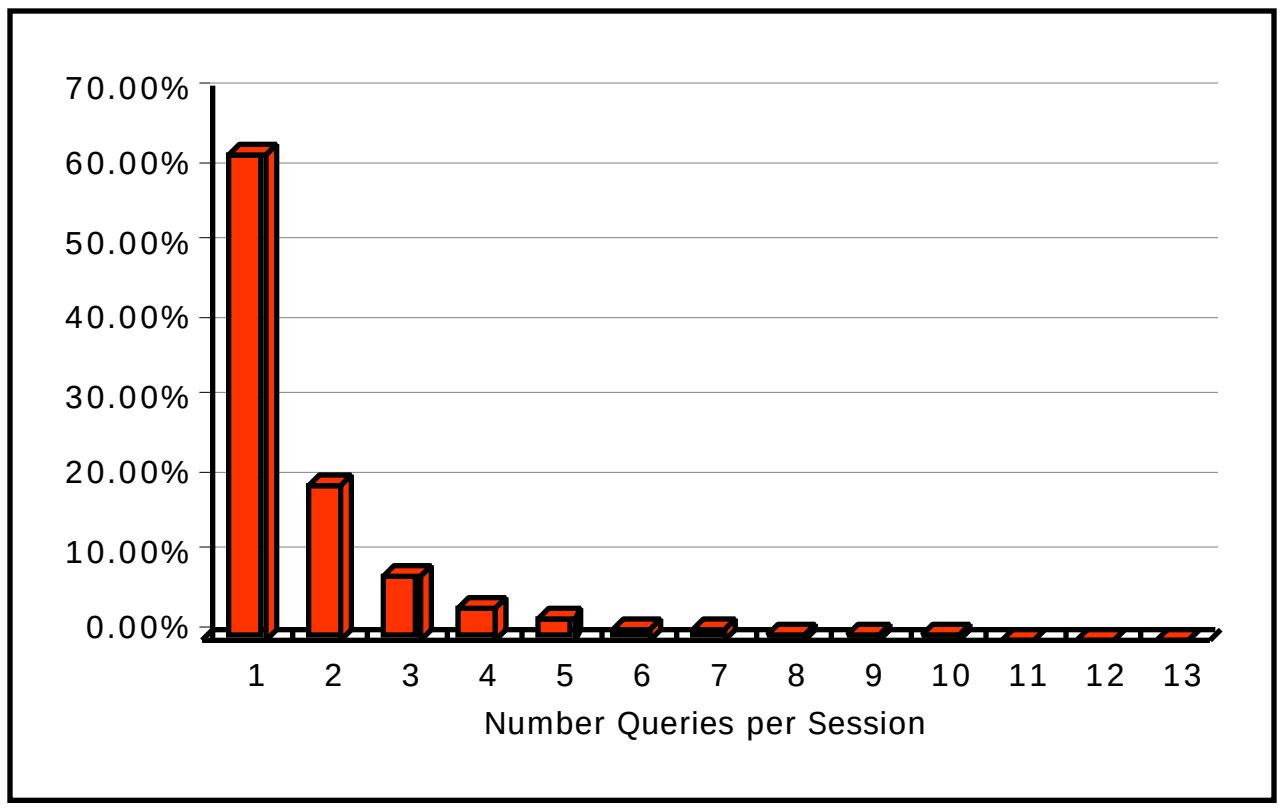

**Figure 12: Distribution of queries per MARIAN session during use study.**

MARIAN query fields had an average of 2.55 terms and a mode of 2, at a time when general-purpose Web queries (at Virginia Tech as elsewhere) were averaging scarcely more than one term per query. We are unsure of the significance of this. Is it because library users were more sophisticated searchers? Because MARIAN produced more understandable results from multi-word queries than the Web search engines of the day? Or just because we gave (and continue to give) users a larger space to enter their query text? We don't know, but we suspect all of the above.

Users made wide use of MARIAN's multiple-access facility. This is significant for any digital library system, but particularly in the context of library catalog data. Library catalog records are complex data structures. MARIAN handles full MARC records [23], but makes a radical reduction into a set of categories more reminiscent of the Dublin Core [8]. Specifically, we assign any fields and subfields that we believe have more content than noise to one of six categories:

> Personal Name, Corporate Name, Conference Name
> Title, Subject, Note

Each text field in a query form object can be configured to cover one or more categories. In 1996, several interfaces were in use. All had four configurable fields, including one preset to "Author" (personal, corporate or conference), one to "Title", and one to "Any Part." The most common interface, that provided by the current Web gateway, had its remaining field preset to "Subject." Close scrutiny of the log files reveals that some use was still being made of older interfaces, where the fourth was preset to "Title, Subject."

Most queries made use of a single field covering a single category (Fig. 13). However, users did combine more than one query field in a significant number of cases. They also made significant use of expanded coverage, using the "Title, Subject" field even when it was not a preset, and using "Title, Subject, Notes" as well. "Any part" was also wisely used, both as a single field and in combination.

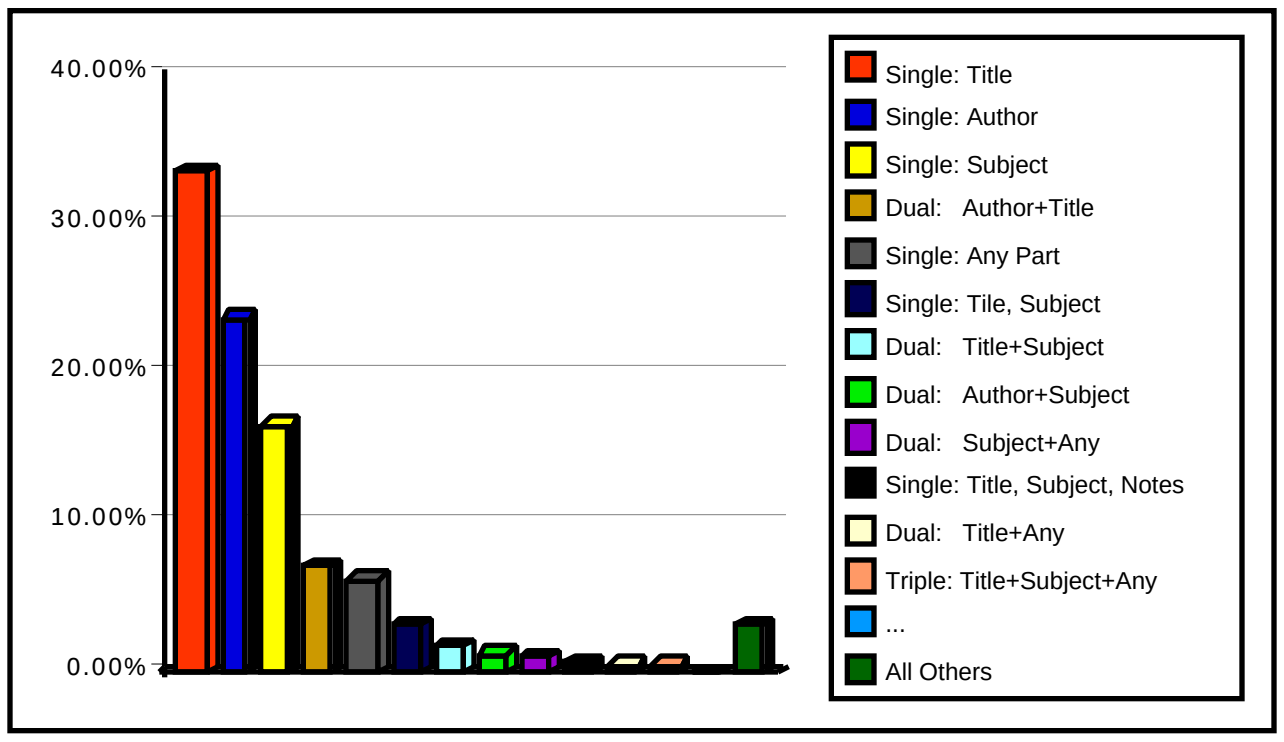

**Figure 13: Most common fields and field combinations during the MARIAN use study. Note use of both multiple text fields (denoted with plus signs) and fields covering multiple categories (denoted with commas).**

Very rarely (2.2%) did users use three query fields, and almost never (< 0.2%) all four fields provided. In fact, most of the four-field queries consisted of the same query terms copied into all four preset fields, an information need statement not obviously different from a single "Any part" query. This unexpected usage encouraged us to redesign the query page instructions; the lack of use for more than two fields encouraged us to simplify the query form.

Another unexpected usage was more worrisome: a significant number of queries were encountered with unrestricted text in the "Author" field (e.g., "Author: drunk driving" or "Author: journal of broadcasting"). We theorize that users were not trying to restrict these searches to works with those words in an author name, but rather were simply not attending to the field preset. The query form page displays the "Author(s)" field first, as is common in bibliographic presentations. Users with small monitors, however, may have been unable to see the other fields. Most Web searchers of the day, as now, presented only a single editable field in a query form: our supposition was that MARIAN users with small browser windows, seeing only the "Author(s)" field, ignored the label, entered text, and hit the "Do Search" button.

## MODIFYING FOR USE

The first order of business in revising the MARIAN WebGate pages was thus clearly the query form. We needed a form that would fit into a browser running on a common 13" (640x480) monitor. We wanted to make instructions clearer and more obvious. In addition, usability testing of the related Envision system [17] had shown that examples could be integrated into query forms without distracting users, and that the users found such examples useful. So we wanted to add examples. Finally we wanted to create a more integrated page, where users could confidently tell that they had the entire form on screen.

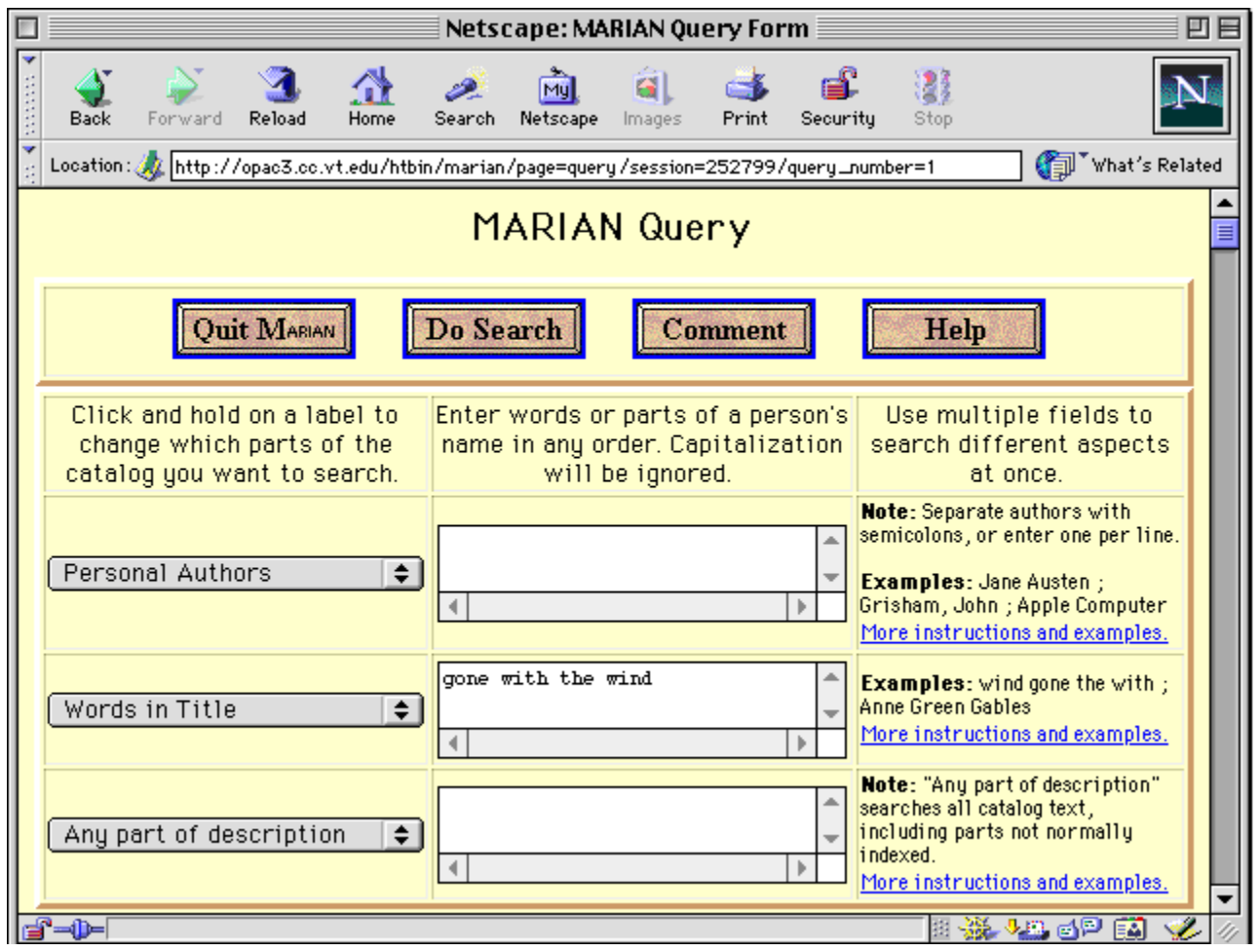

**Figure 14: Web query form object after use study and subsequent redesign, as displayed on a 640x480-pixel screen.**

Fig. 14 shows the redesigned query form. The fields have been reduced to three, examples and instructions integrated, and the form layout generally tightened up. In addition, concurrent improvements to the back end had made possible a choice of author settings, including personal, corporate, and conference authors, as well as any combinations of the categories. The top field can thus be preset to "Personal author," a faster search than generalized author, and demonstrably more used. Other author categories are available in the pop-up. The other two fields are preset to "Title" and "Any part," although "Subject" was more common in the use study than "Any part." We hoped that by providing one broad coverage in the presets we would keep users aware of the possibilities of both broad and narrow searches. Future use studies may tell us whether that is true.

The appearance of the MARIAN Web pages was thoroughly revised. A uniform light bisque was chosen as a background for the pages, and a uniform design was used for all buttons. The order of the buttons was also standardized, while their presentation was unified with an HTML "table" construct. Buttons for "Help," "Comment," and "Quit" were included on every page, as were buttons to enable common transitions (e.g., "Edit Query" versus "New Query"). In cases like the result set object, where the content of the page was likely to go beyond the 13" screen to which we were designing, buttons were duplicated on the top and bottom of the page.

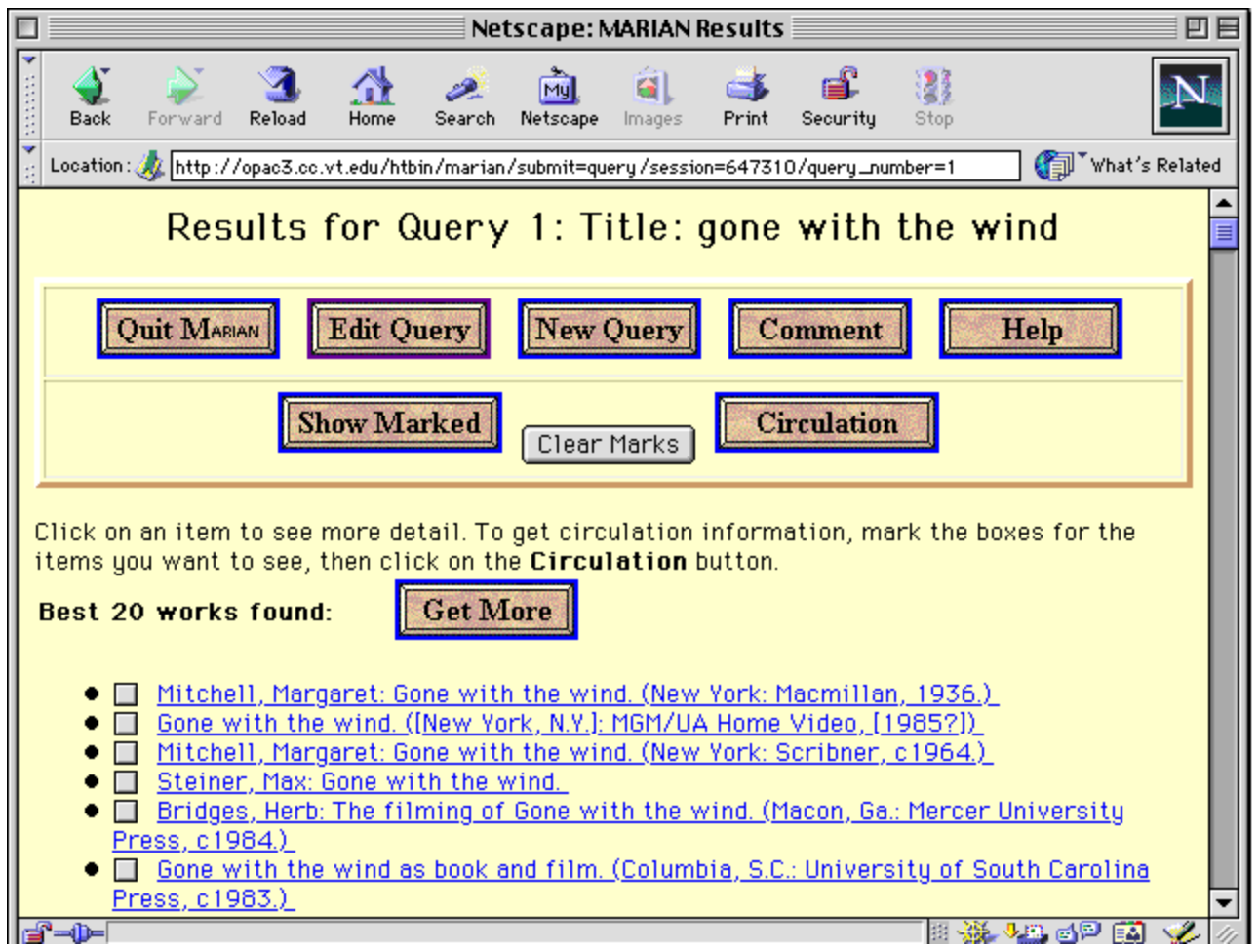

**Figure 15: Web-based result set object with expanded action set presented as buttons.**

User behavior with result sets motivated us to add a "Show Marked" feature to result set objects, so that users could examine several documents at the same time (Fig. 15). Implementing this involved a change that users had also asked for through MARIAN's "Comment" facility: persistent markings in result sets. Like so many user actions, marking an "Of interest" checkbox on the result set page is a local operation within the browser: HTTP only reports the state of checkboxes on a "post" operation. Thus making checkbox state persistent required us not only to add data structures within WebGate, but to modify the result set page so that all operations involved posts. Once that was accomplished, however, the checkboxes became useful for creating subsets and printed bibliographies, as well as informing requests for bibliographic data and circulation information.

All these improvements were implemented within the confines of HTML 2.0. In the summer of 1997, HTML 3.0 was available, but had not penetrated far into our user community. In particular, HTML frames were becoming common on the Web, but only the most up-to-date browsers could handle them. We believe it is central to our mission as a library system to use simple, well-tested, and widely supported constructs only, since digital library systems must serve the full users community, not just those with the newest software.

## CURRENT AND FUTURE DIRECTIONS

Development of MARIAN at Virginia Tech continues. In 1998, two changes were immediately apparent in the WebGate interface. First, WebGate session objects were extended to cache old queries and result sets, making it possible for the user to move back and forth among conversational paths as had been the case in windowing interfaces. Second, the presentation of bibliographic results was enhanced with HTML hyperlinks for each author, subject, and title. This enhancement presents usability problems, which we plan to test.

In each case, the effect of activating a link was to initiate a new search for works with related authors, subjects, and titles. Since the default MARIAN search remains an approximate match, these produce result sets that include not only works with exactly the same author, subject, or title, but also works with similar attributes. In the case of titles, we believe that this is clearly the correct effect. Equally clearly, it is almost always the wrong effect for personal authors: for a personal name, a match of two out of three name components is generally no match at all. In the case of corporate names, conference names and subjects, the effect is less clear. A near match between two textual representations of subjects sometimes denotes similar concepts, other times not. Our next usability evaluation will study what users expect from such links, and how we can best satisfy them.

The MARIAN system is but one component of a digital library. As part of continuing digital library research at Virginia Tech, MARIAN development is undergoing a process of expansion and integration into broader digital library systems. Out first effort in the direction was the Envision system [12, 18] which featured the MARIAN search system running on a mixed collection of bibliographic records and full-text documents, coupled to a seminal visualization interface. Current efforts include personalization and integration into the surrounding digital library framework.

Personalization means that when a users identify themselves to the system, the system can provide an interface customized according to their preferences. We believe that with a personalized interface, a system can serve more users' needs and serve users' needs better. Currently, the MARIAN system supports several personalization features. The number of documents which will be returned by the system each time the user presses the "Get More" button can be configured for each query. The time to wait for the result documents to return can be configured globally or for each query. This is helpful since sometimes it may take a long time for documents to be returned due to network traffic. Display styles for documents and bibliographic records, as well as sorting principles for result sets, will be configurable in the next release of the MARIAN WebGate. We are also considering giving users control of the look and feel of MARIAN pages, including colors and page styles.

The next release will also support a global history over all sessions for a single user. This will allow the user to see any query performed in the past along with the results retrieved at that time. The user can easily perform the query against the current collection or some other collection, or modify and then perform it. Usability issues raised by this feature include organization of the global history, whether users can delete queries from the history, and how to mark subsets or conversational paths in the history. Clearly, much work still needs to be done.

## CONCLUSION

Libraries have come to exist in response to needs in human communities. Digital libraries are no exception. Digital library components – tools, frameworks, and collections – may grow out of a conception by providers, but digital libraries as a whole will thrive or wither only as they serve or fail to serve their user communities. Our experience with MARIAN serves to validate this proposition. We have found that user behavior is not always predictable, but that tracking it carefully can contribute to a usable product. And we have found that it matters not at all how good a search system you have if the users cannot find the "Do Search" button.

This paper presents the evolution of a digital library component, from a provider conception to a well-subscribed service. That process was mediated by attention to users and user studies and by two formal instruments: usability evaluation and statistical log analysis. Both instruments aided in design because they provided insight into the mental models and overt behavior of users. As we progress in the evolution of full digital library systems, we expect our development to continue to be mediated by an emphasis on users, use and usability.